\documentclass{article}
\usepackage{INTERSPEECH2018,amsmath,graphicx,url, float}

\title{Efficient keyword spotting using time delay neural networks}
\name{Samuel Myer, Vikrant Singh Tomar}
\address{Fluent.ai Inc., Montreal, Canada}
\email{sam.myer@fluent.ai, vikrant@fluent.ai}

\begin{document}
%\ninept
%
\maketitle
\begin{abstract}
This paper describes a novel method of live keyword spotting using a two-stage time delay neural network. The model is trained using transfer learning: initial training with phone targets from a large speech corpus is followed by training with keyword targets from a smaller data set.  The accuracy of the system is evaluated on two separate tasks.  The first is the freely available Google Speech Commands dataset. The second is an in-house task specifically developed for keyword spotting.  The results show significant improvements in false accept and false reject rates in both clean and noisy environments when compared with previously known techniques.  Furthermore, we investigate various techniques to reduce computation in terms of multiplications per second of audio.  Compared to recently published work, the proposed system provides up to 89\% savings on computational complexity.
\end{abstract}


\section{Introduction}
\label{sec:intro}

Keyword spotting is an essential feature in modern hands-free voice control devices, where the user speaks a predefined keyword to ``wake-up'' the device before speaking a complete command or query to the device. This keyword is also referred to as a ``wake-word''. Unlike large vocabulary speech recognition systems, keyword spotting algorithms use a simpler model that only detects whether a phrase or small set of phrases are spoken. Once a wake-word has been detected, then a large vocabulary model could be used to decode the user query that might follow.

Much research has been done in recent years on improving keyword spotting methods. Recent works have suggested use of fully connected neural networks \cite{Chen14}, convolution neural networks (CNNs) \cite{Sainath15,Zhang17}, and recurrent neural networks (RNNs)\cite{Lengerich16, Audhkhasi17}. RNNs have also been combined with convolutional layers \cite{Arik17}. Recently, two dimensional Grid-LSTM RNNs capable of learning sequences in both the time and frequency dimensions have also been shown to produce good results albeit with higher computational complexity \cite{Li17}.

In this work, we present a novel two-stage time-delay neural network architecture (TDNNs) for keyword spotting. We also discuss various optimizations for efficient implementation of the presented system. In a TDNN, different layers or sets of layers can act on different time scales \cite{Waibel89}. As such, it can be seen as a type of CNN \cite{Lecun95} operating over the time dimension. In a TDNN model, the first few layers look at smaller time scales and produce more abstract higher level features. The later layers take larger time windows over these abstract features as input. During training and inference, the sequence of input features are repeatedly shifted in time and fed to the model, producing another sequence as output. This architecture reduces the amount of computation required, as compared with a fully connected network.

This paper explores several improvements over existing systems.  There exists some recent work where the authors have used TDNNs for keyword spotting in combination with a hidden Markov model (HMM) \cite{Sun17}.  However, the presented method uses an end-to-end TDNN architecture and avoids the need for a separate HMM model. The system consists of two sets of layers. The first set is trained with phone label targets as an intermediate representation and the second set learns to predict the keyword targets.  The first set of layers is initialized independently of the second set with weights trned on a large vocabulary task. Transfer learning is then used to train the second sets of layers on top of the first set. Furthermore, we investigate using frame skipping and caching to reduce computation in a real-time scenario. 

\section{System Description}
\label{sec:model}

An effective keyword detection system must minimize both false positives and false negatives to provide an acceptable user experience. The number of operations and memory usage should also be kept low to reduce power drain and conform to hardware limitations. This section presents variations of our TDNN architecture and measures them on these performance objectives to arrive at an optimal configuration.

\subsection{Architecture}
\label{ssec:architecture}

\begin{figure}[htb]
\begin{minipage}[b]{1.1\linewidth}
  \centering
  \centerline{\includegraphics[width=8.5cm]{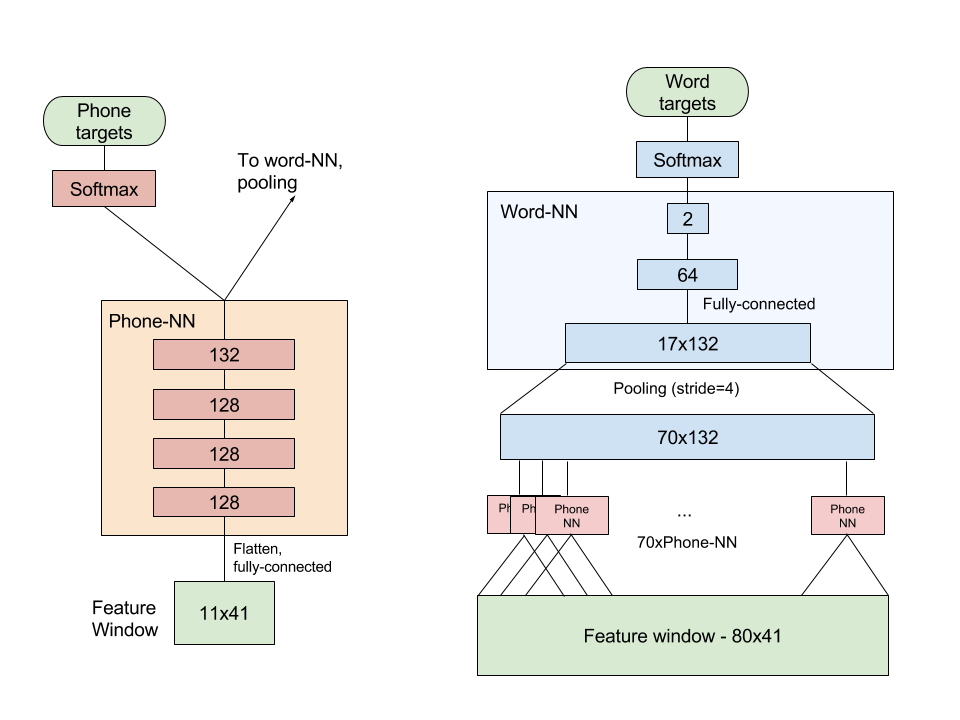}}
\end{minipage}
\caption{Architecture of the proposed model.  The left side shows a zoom-in of the phone-NN layers.  The right side shows the complete architecture.}
\label{fig:tdnn}
\end{figure}

The model used in this paper is shown in Figure \ref{fig:tdnn}. The number of parameters per layer are provided in Table \ref{tab:arch}. The model consists of two sets of layers which can be seen as two separate neural networks. All our experiments use 41-dimensional log-Mel filterbank (FBANK) features.  These features are extracted from the input speech using a frame size of 25 ms and frame shift of 10 ms. The FBANK features are normalized so that they have approximately zero mean and unit variance. The first set of layers takes FBANK features as input and are trained with phone targets as an intermediate representation. For easy reference, we call this set the phone-NN. This set of layers take a context of 5 frames in the past and 5 frames in the future for a total context of 11 frames, or 125 ms.  The phone-NN is shown in red in Figure \ref{tab:arch}.

The phone-NN is trained on monophone targets instead of triphones or senones.  However, state-dependent variants of phones are used.  There are 3 variants for each non-silence phone, and 5 variants for each silence type (3 types of silence were used).  This configuration is motivated by Kaldi\cite{kaldi} and results in a model with higher discriminative capacity without the overhead of a larger output layer, as would occur when using triphones.

The output of the phone-NN layers, without the softmax layer, is used as input for the second set of layers. These layers are trained with the keyword labels as the training targets, therefore we refer to this set as the word-NN. The word-NN is shown in blue in Figure \ref{tab:arch}.  To reduce the input dimensionality, the phone posteriors are first max-pooled along the time axis with a pooling size of 5 frames and a stride of 4. The pooled phone posteriors are passed to two fully connected followed by a softmax layer. 

Altogether, the whole network looks at a input context of 80 frames covering 815 ms of speech, consisting of 69 frames in the past and 10 frames in the future. The output layer has one unit for each target keyword and one unit for background/filler speech. The input window is shifted in time across the FBANK features producing a sequence of keyword probabilities. 

During decoding, the keyword probabilities are smoothed using a moving average filter with a width of 9 frames.  We applied smoothing in a similar manner to the method described in the paper \cite{Chen14,Sainath15}.   A threshold is applied to the smoothed probabilities, and keyword detection is triggered when one the keyword probabilities goes above the threshold.

\begin{table}
\begin{center}
\begin{tabular}{|c|c|c|c| } 
\hline
Layer & Inputs & Outputs & \# Weights \\ 
\hline
phone-1 & 451 & 128 & 57728 \\ 
phone-2 & 128 & 128 & 16384 \\ 
phone-3 & 128 & 128 & 16384 \\ 
phone-4 & 128 & 132 & 16896 \\ 
word-1 & 2244 & 64 & 143616 \\ 
word-2 & 64 & 2 & 128 \\ 
\hline
Total & & & 251136 \\ 
\hline
\end{tabular}
\caption{TDNN parameters per layer}
\label{tab:arch}
\end{center}
\end{table}
 
\subsection{Computational complexity}
\label{ssec:multiplies}

Table \ref{tab:power} presents the computational complexity of the models discussed in this paper, measured as the number of multiplications necessary per second of input audio. In a real-time system, this metric is roughly proportional to power consumption, which is relevant in environments with limited hardware capabilities.  Two methods are described here to reduce multiplications per second: caching intermediate results and frame skipping.

The phone-NN only looks at small patches of the input data. Recalculating all of these patches whenever the full TDNN is shifted a time step results in a lot of redundant computation. The amount of computation can be greatly reduced using dynamic programming. The output phone posteriors from the first set of layers is cached in a buffer. Then, only the rightmost patch at each level of the TDNN needs to be calculated at each time step.

Another way of reducing computation is by skipping frames during inference. Since the region of interest in the input, where the keyword is spoken, might span several frames, it is reasonable to argue that the network might still be able to capture sufficient relevant information even if some frames are skipped during decoding. This is discussed in more details in Section \ref{ssec:result_frameskipping}.

\section{Experimental Setup}
\label{sec:experiment}

The proposed model is evaluated on two separate speech tasks: one is an in-house dataset specifically developed for keyword spotting tasks and the other is a freely available dataset called Speech Commands \cite{GSC}. 

The primary motivation for using the in-house dataset is to test the system in a wide variety of background speech and acoustic conditions. The dataset consists of long speech recordings created by concatenating the keyword (``Fluent"), short pauses, non-keyword filler speech, and noise in random order. 

The amplitude of the keyword speech and filler speech is randomly varied to simulate speakers with varying loudness. For noisy conditions, simulated speech-in-noise data is created by mixing three different noise types, namely babble, music and street, with clean speech at an average of 10 dB SNR. The test set is created in a similar manner, but using different noise files than those used to create the training set. Furthermore, long conversational speech that does not contain any examples of the keyword is used to train the models for background or out-of-vocabulary speech. The resulting dataset consists of 50 hours of training data with 5913 repetitions of the keyword, and 22 hours of testing data with 1563 repetitions of the keyword. The results are evaluated on false alarms per hour and false reject rate.   

Most of the previous work in keyword spotting research has used internal datasets that make it difficult for general scientific community to compare performance of different models. We realize this is also the case with our in-house dataset. To avoid unreproducible research, we perform a second set of experiments on the freely available Speech Commands dataset \cite{GSC}. For these experiments, we compare our work with that reported in \cite{Zhang17}, where the authors have used the same dataset. This dataset contains 64752 clean recordings of 30 commands. While the in-house task described above is designed to recognize only one keyword, in this task we are recognizing 10 keywords concurrently. This is done to create experiments in line with the work reported in \cite{Zhang17}. The 10 commands that are used as target keywords are ``Yes", ``No", ``Up", ``Down", ``Left", ``Right", ``On", ``Off", ``Stop" and ``Go". The remaining 20 commands are used as filler words: ``Bed", ``Bird", ``Cat", ``Dog", ``Happy", ``House", ``Marvin", ``Sheila", ``Tree", ``Wow", and numbers zero through nine.

The original Speech Commands dataset consists of individual commands. However, a keyword spotting system used in realistic scenario is required to identify the target keywords from a continuous stream of incoming speech. In order to simulate this behavior, we also created a derivative dataset from the Speech Commands dataset by concatenating multiple commands together at varying amplitude level. The order of the commands is randomized before concatenation to further simulate real world behavior. Experiments on the Speech Commands dataset are performed on both the original as well as the derivative dataset.  For this experiment, accuracy is measured as the percentage of utterances that are correctly classified as either the appropriate keyword or filler.  This metric is chosen to match baseline results found in the literature \cite{Zhang17}.

\subsection{Transfer learning}
\label{ssec:transfer}

Transfer learning is a method for initializing weights by first training the network on a larger corpus for a related task \cite{Pratt93} and then using some of the layers of this network to train on the main task.  This allows the network to build upon the learning from the larger amount of data of the related task and is particularly useful for scenarios where only a limited amount of data may be available for the main task.  Transfer learning and multi-task learning \cite{Sainath15, Panch16, Sun17} are common practices in keyword spotting tasks, where the amount of training data available is often limited. Transfer learning also helps reduce overfitting \cite{Panch16}. 

In the network architecture used in this work, a softmax layer is added after the phone-NN.  The phone weights are trained on a large vocabulary continuous speech recognition (LVCSR) corpus using 132 phone targets. Next, the softmax layer is removed and  word-NN layers are added on top of the phone-NN. The entire network is then trained using the keyword dataset.

To ensure a fair comparison,  transfer learning is applied to all the experiment presented in Section \ref{sec:results}. This include the baseline models as well as our proposed model. All the models are initialized using transfer learning on the same LVCSR task.

\section{Results}
\label{sec:results}
This section gives a brief description and results of three experiments: (i) comparison against Speech Commands dataset baseline, (ii) performance on in-house dataset, and (iii) computational savings achieved by using frame skipping without compromising accuracy. The \textit{cnn-one-fstride4} keyword spotting system described in \cite{Sainath15} is used as the baseline architecture in this work for both the Speech Command dataset as well as the in-house dataset.  The \textit{cnn-one-fstride4} model consists of a convolutional layer strided in the frequency dimension, followed by three fully connected layers.

\subsection{Results on the in-house dataset}

\begin{figure}[t]
	\begin{minipage}[b]{\linewidth}
		\centering
		\centerline{\includegraphics[width=6.25cm]{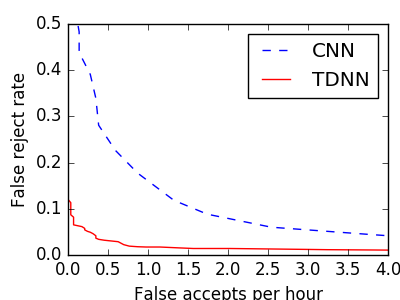}}
		%  \vspace{1.5cm}
		\centerline{(a) Clean}\medskip
	\end{minipage}
	\hfill
	\begin{minipage}[b]{\linewidth}
		\centering
		\centerline{\includegraphics[width=6.25cm]{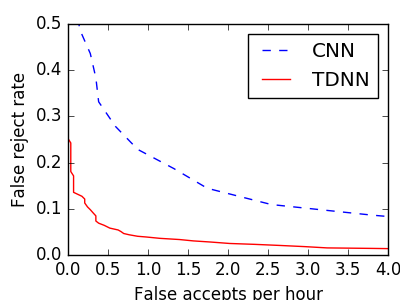}}
		\centerline{(b) Noisy - 10dB SNR}\medskip
	\end{minipage}
	\caption{ROC curves for the presented TDNN method vs the CNN baseline on the in-house dataset.}
	\label{fig:res_baseline}
\end{figure}

Table \ref{tab:power} provides a summary of each of the models discussed.  The second and third columns of the table list the number of parameters and multiplications per second performed during inference for each model respectively.  The fourth and fifth columns present the experimentally determined false rejection rates (FRR) for each model on clean and noisy data respectively.  All false rejection rates in this section are reported for a fixed false alarm rate of 0.5 per hour. The row labeled ``CNN" provides the numbers corresponding to the baseline \textit{cnn-one-fstride4} model. The row labeled ``TDNN'' provides numbers corresponding to the proposed model. The other two rows in the table are described in Section \ref{ssec:result_frameskipping}. It can be seen from the table that the proposed TDNN network results in a lower false reject rate --- 87\% lower on clean data and 71\% lower on noisy data --- relative to the baseline CNN model.  Furthermore, the receiver operator characteristic (ROC) curves in Figure \ref{fig:res_baseline} shows that the TDNN consistently performs better when the activation threshold is varied.  This is true for both clean and noisy data with 10dB SNR.

\begin{table}
	\begin{center}
		\begin{tabular}{|c||c|c|c|c| } 
			\hline
			Model & Params & Mul / s & FRR & FRR \\ 
			& & & clean & noisy \\ 
			\hline
			CNN \cite{Sainath15} & \textbf{122K} & 50.3M & 22.7 & 27.9 \\ 
            Large CNN & 193K & 114.5 M & 7.6 & 19.9 \\
			TDNN & 251K & 25.1M & \textbf{3.1} & \textbf{5.8} \\ 
			TDNN-skip2 & 251K & 12.6M & 4.8 & 7.6 \\ 
			TDNN-skip4 & 251K & \textbf{6.28M} & 3.9 & 8.4 \\ 
			\hline
		\end{tabular}
        \vspace{1em}
		\caption{Results for in-house dataset. For each model, the table shows the number of parameters, multiplications per second, and false reject rate in percent on clean data (FRR clean) and 10 dB SNR noisy data (FRR noisy). FRR values are for a false alarm rate of 0.5 FA/hr.}
		\label{tab:power}
	\end{center}
\end{table}

The large gap in accuracy between the TDNN and \textit{cnn-one-fstride4} model can be partially explained by the difference in input window size ---  815 ms for the TDNN versus 335 ms for the \textit{cnn-one-fstride4} model.  To test the effect of input window size on accuracy, a larger CNN with input window size of 815 ms (80 input frames) was trained.  Aside from the input size, the rest of the CNN had the same architecture as \textit{cnn-one-fstride4}, convolving in the frequency dimension only.  The CNN with increased frame size had improved accuracy (7.6\% FRR in clean and 19.9\% in noisy environment), but still was less accurate than the TDNN model.  To ensure that the difference in accuracy is indeed due to input window size and not number of parameters, another CNN was trained with the same number of parameters as the TDNN. However, increasing parameters without increasing input size did not improve accuracy.

\subsection{Results on Speech Command dataset}
\label{ssec:result_baseline}

\begin{table}[t]
	\begin{center}
		\begin{tabular}{|c|c|c| } 
			\hline
			Model & Error \% & Error \% \\ 
			& (original) & (derivative) \\ 
			\hline
			CNN (from literature) & 15.4 \cite{Zhang17} & n/a \\ 
			CNN & 6.5 & 24.8 \\ 
			TDNN (ours) & \textbf{5.7} & 15.2 \\ 
			\hline
		\end{tabular}
        \vspace{1em}
		\caption{Results on Speech Commands dataset. Percentage of commands correctly labeled on 10-keyword Speech Commands dataset.  Our TDNN is compared against the \textit{cnn-one-fstride4} model from \cite{Sainath15}. Two numbers are given for the CNN model: the first comes from the literature \cite{Zhang17} and the second was measured by us.  Accuracy is measured on the original data, as well as derivative data, as described in \ref{sec:experiment}. }
		\label{tab:baseline}
	\end{center}
\end{table}

The results for the speech commands dataset are presented in Table \ref{tab:baseline}. The accuracy is presented along with that reported in \cite{Zhang17} on the Speech Command dataset and the proposed model as described in Section \ref{sec:model}. The table contains results reported in \cite{Zhang17} for the baseline CNN, and the results we obtained for the baseline CNN and the TDNN proposed in this work. The results are presented for the original Speech Command dataset as well as our derivate dataset. It should be noted that the baseline results that we obtained in this work are higher than those reported in \cite{Zhang17}. This is perhaps due to difference in optimizations such as posterior smoothing \cite{Sainath15}, which improves accuracy on the baseline model. When comparing the presented model to the baseline in Table \ref{tab:baseline}, the presented model shows a relative error reduction of 12 \% on the original data.  On the derivative data, where the audios were stitched together with varying amplitude to better simulate real-world conditions, error decreased by 39 \%.

\subsection{Frame skipping}
\label{ssec:result_frameskipping}

\begin{figure}[t]
\begin{minipage}[b]{\linewidth}
  \centering
  \centerline{\includegraphics[width=6.25cm]{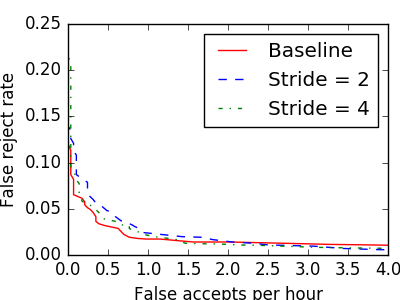}}
%  \vspace{1.5cm}
  \centerline{(a) Clean}\medskip
\end{minipage}
\hfill
\begin{minipage}[b]{\linewidth}
  \centering
  \centerline{\includegraphics[width=6.25cm]{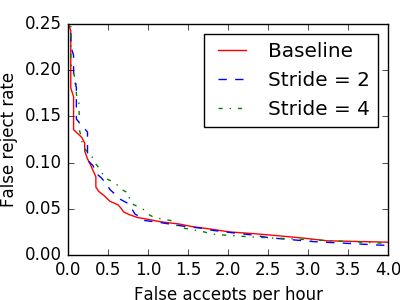}}
  \centerline{(b) Noisy - 10dB SNR}\medskip
\end{minipage}
\caption{ROC curves with frame skipping}
\label{fig:res_frameskip}
\end{figure}

\begin{figure}[H]
	\centering
	\includegraphics[width=8.5cm]{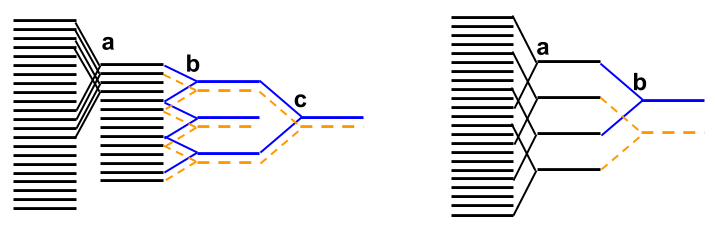}
	\caption{Illustration of frame skipping mechanism.  Blue lines represent the first output frame, and orange lines represent the second output frame.  \textbf{Left}, without frame skipping: a) Phone features are calculated at every time step  b) Phone features are pooled to reduce dimensionality  c) Keyword features are calculated at every time step.  \textbf{Right}, with frame skipping: a) Phone features are calculated with a stride of 4 time steps, without pooling  b) Keyword features are calculated at every fourth time step. }
	\label{fig:frame_skipping}
\end{figure}

As mentioned in Section \ref{ssec:multiplies}, this work also uses frame-skipping to further reduce computation without significantly diminishing accuracy. Experiments were done with strides of 2 and 4 for both stages of the TDNN, as illustrated in figure \ref{fig:frame_skipping}. ROC curves for these experiments are given in Figure \ref{fig:res_frameskip}. It can be seen from the ROC curves that the impact of frame-skipping on accuracy of keyword spotting is very minimal. Resulting FRRs are 3.1\% without frame skipping, 4.8\% with a stride of 2, and 3.9\% using a stride of 4. This indicates that frame skipping is a good way to reduce computation without greatly impacting accuracy.  The frame skipping methods here could be combined with other optimization methods suggested in the literature such as quantization \cite{Lei13,Zhang17} or binarization \cite{Xiang17, Hubara16} to further improve performance.

\section{Conclusions}
\label{sec:conclusions}
This paper has presented a novel method of using time delay neural network for keyword spotting. 
The network weights were initialized using transfer learning on a related LVCSR task. The proposed TDNN architecture was compared against the \textit{cnn-one-fstride4} model described in \cite{Sainath15}. It has been shown that with a false alarm rate of 0.5 FA/hr, the presented TDNN model significantly reduces the false reject rate on realistic data compared to the baseline CNN model, while reducing the number of multiplications by 50\%.  However, on the Speech Commands dataset without any modifications, accuracy improvement was more limited.  Furthermore, it has been shown that the number of multiplications can be drastically reduced by applying frame skipping without significantly affecting accuracy.  In a continuously listening system, this allows for acceptable accuracy while running on low-power devices.

%\vfill
%\pagebreak

\bibliographystyle{IEEEbib}
\bibliography{strings.bib}

\end{document}